\documentstyle[12pt]{article}   
\newtheorem{proposition}{Proposition}
\newtheorem{theorem}{Theorem}
\newcommand{\V}
   {\raisebox{.3ex}{$\scriptstyle\bf\backslash$}\!\mbox{\rm V}}

\newcommand{\VH}{\mbox{\bf H}}
\newcommand{\R}{\mbox{\bf R}}
\newcommand{\T}{\mbox{\bf T}}
\newcommand{\W}{\mbox{\bf W}}
\newcommand{\0}{\mbox{\bf O}}
\newcommand{\fracc}[2]{\frac{\textstyle #1}{\textstyle #2}}

\newcommand{\nul}{\raisebox{2ex}{\null}}
\newcommand{\unity}{\mbox{\bf I}}
\begin{document}

\title{\centerline{\bf The trace formulas  yield the inverse metric formula}\author{Ronaldo Rodrigues Silva\thanks{E-mail: ronaldor{@}cat.cbpf.br}
 \\{\small Centro Brasileiro de Pesquisas F\'{\i}sicas}\\
 {\small Rua Dr.\ Xavier Sigaud 150, 22290-180, Rio de Janeiro, RJ, Brazil} }}
\date{}
\maketitle
\begin{abstract}
\indent\par
It is a well-known fact that the first and last 
non-trivial coefficients of the characteristic 
polynomial of a linear operator are respectively its 
trace and its determinant.  This work shows 
how to compute recursively all the coefficients 
as polynomial functions in the traces 
of the successive powers of the operator.  
With the aid of Cayley-Hamilton's theorem 
the trace formulas provide 
a rational formula for the resolvent kernel and 
an operator-valued null identity for each finite dimension 
of the underlying vector space. The 4-dimensional resolvent formula allows an algebraic solution 
of the inverse metric problem in general relativity.  

\end{abstract} 

\section{Introduction}


\indent\par
The eigenvalue problem arises in a variety of different 
branches of mathematical physics.  
For instance, it is well-known that 
quantum systems reach stationary states 
that are the eigenvectors of 
a suitable linear operator defined 
throughout a complex vector space 
representing the physical quantum states.  

In the infinite dimensional case 
real and complex analytical methods have been developed 
to compute the eigenvalues of a linear operator.  
We mention, for instance, global variational methods, 
which consist of investigating stationary levels 
of a suitable energy functional, and local perturbative methods 
by means of complex analytic continuation.  

On the other hand, 
if the underlying vector space is finite dimensional
the eigenvalue problem becomes 
an algebraically well-posed problem 
by means of the characteristic polynomial.  
It is exactly this feature that justifies 
the algebraic approach in this work.  
Following this general device, 
the algebraic environment is set by the complex number field. 
After working out a few tools, 
the most important of these being Newton's identities, 
and setting very fundamental concepts 
and basic results in complex linear algebra, 
we succeed in achieving the mathematical core of this work: 
a recursive algorithm to compute 
the coefficients of the characteristic polynomial 
as algebraic functions in the traces of the successive powers 
of the linear operator.  

With the aid of Cayley-Hamilton's theorem 
the trace formulas already obtained provide 
an operator-valued null identity for each finite dimension 
of the underlying vector space.  As a by-product 
it sheds light on the algebraic structure 
of the associative algebra of complex linear operators.  
The computational skill just attained is enough to yield 
an operator-valued polynomial with rational coefficients for the  
finite-dimensional resolvent kernel, which improves a known result by revealing 
its  rational dependence 
with respect to the  spectral variable, as well as with respect to the linear operator.  

In the context of general relativity,
the 4-dimensional characteristic formula endows a polynomial expression for the volume scalar density. Furthermore, the 4-dimensional resolvent formula yields 
a tensor-valued  third-degree polynomial 
for the inverse metric, thus avoiding the computacional and formal drawbacks of the Neumann series.        

\section{Newton's identities}
\protect\label{Identities}

\indent\par
The fundamental theorem of algebra \cite{algebra} 
together with Euclid's division algorithm imply 
algebraic closureness \cite{Lang} of the complex number field; 
this basic result is encompassed without proof by 
\begin{theorem}
\protect\label{Algebra}
If $\,p(z)=z^n+D_1z^{n-1}+\cdots+D_{n-1}z+D_n$ 
is a polynomial with complex coefficients $\,D_k$ 
then there exist complex numbers 
$\,\lambda_1,\,\lambda_2,\,\ldots,\,\lambda_n$, 
called roots of $\,p$, such that 
$\,p(z)=(z-\lambda_1)(z-\lambda_2)\cdots(z-\lambda_n)$.  
\end{theorem} 

As a consequence of the identity principle, 
the following relations between coefficients and roots hold:
\begin{equation}
\protect\label{lambda}
\begin{array}{r@{\hspace{.3em}}l}
-D_1&=\lambda_1+\lambda_2+\cdots+\lambda_n, \\
+D_2&=\lambda_1\lambda_2+\cdots+\lambda_{n-1}\lambda_n, \\
&\hspace*{.2em}\vdots \\
(-)^kD_k&=\sum\mbox{$k$-products of $\lambda$'s}, \\
&\hspace*{.2em}\vdots \\
(-)^nD_n&=\lambda_1\lambda_2\cdots\lambda_n.
\end{array}
\end{equation}
\

The right-hand side of (\ref{lambda}) defines the elementary symmetric 
functions \cite{symmetric} in the variables 
$\,\lambda_1,\,\lambda_2,\,\ldots,\,\lambda_n$.  
Next to them, the most important symmetric functions 
are the sums of like powers:
\begin{equation}
\protect\label{Tk}
\begin{array}{l}
T_1=\lambda_1+\lambda_2+\cdots+\lambda_n, \\
T_2=\lambda_1\mbox{}^2+\lambda_2\mbox{}^2+\cdots+\lambda_n\mbox{}^2, \\
\hspace*{5em}\vdots \\
T_k=\lambda_1\mbox{}^k+\lambda_2\mbox{}^k+\cdots+\lambda_n\mbox{}^k, \\
\hspace*{5em}\vdots 
\end{array}
\end{equation}

$\,D_k$ and $\,T_k$, besides being symmetric, 
are homogeneous functions of degree $k$ in the variables 
$\,\lambda_1,\,\lambda_2,\,\ldots,\,\lambda_n$.  We shall derive 
a set of recursive relations connecting $\,D$'s and $\,T$'s by which 
the $\,\lambda$'s are eliminated from (\ref{lambda}) and (\ref{Tk}).  

To pursuit this goal we need two lemmas.  The first one is 
an improved version of the remainder theorem.
\indent\par
\begin{proposition}
\protect\label{seila}
If $\,p(z)=z^n+D_1\,z^{n-1}+\cdots+D_{n-1}\,z+D_n$ is a polynomial 
with complex coefficients $\,D_k$ and $\,\lambda$ 
is a complex number then 
$$
\begin{array}{l@{\hspace{.3em}}l}
\fracc{p(z)-p(\lambda)}{z-\lambda}=&z^{n-1}+(\lambda+D_1)\,z^{n-2}
+(\lambda^2+D_1\,\lambda+D_2)\,z^{n-3}+\cdots \\
& +\,(\lambda^{n-1}+D_1\,
\lambda^{n-2}+\cdots+D_{n-1}).
\end{array}
$$
{\bf Proof}: By induction on the degree. For n=1,
\,\,$p(z)-p(\lambda)=(z+D_1)-(\lambda+D_1)=z-\lambda.$
For generic $n$, \,\,$p(z)-p(\lambda)=$ 
$$
\begin{array}{l@{\hspace{.2em}}l}
& = (z^{n}+D_1\,z^{n-1}+\cdots+D_{n-1}z+D_n)-
\left(\lambda^{n}+D_1\,\lambda^{n-1}+\cdots
+D_{n-1}\lambda +D_n\right) \\
& =z\left(z^{n-1}+D_1\,z^{n-2}+\cdots+D_{n-1}\right)-
\lambda\left(\lambda^{n-1}+D_1\,\lambda^{n-2}+\cdots
+D_{n-1}\right) \\
& =z\left((z^{n-1}+D_1\,z^{n-2}+
\cdots+D_{n-1})-\left(\lambda^{n-1}+D_1\,\lambda^{n-2}+
\cdots+D_{n-1}\right)\right)+ \\
&\hspace{1em}\,\,+\,(z-\lambda )(\lambda^{n-1}+D_1\,\lambda^{n-2}+
\cdots+D_{n-1}). \\
\end{array}
$$
From the inductive hypothesis the last expression reads
$$
\begin{array}{l@{\hspace{.2em}}l}
&\,\,\,\,\,\,\,z\,(z-\lambda )\,(z^{n-2}+(\lambda+D_1)\,z^{n-3}+\cdots
+(\lambda^{n-2}+D_1\,\lambda^{n-3}+\cdots+D_{n-2}))\,+ \\
& \,\,\,\,\,\,\,+\,(z-\lambda )(\lambda^{n-1}+D_1\,\lambda^{n-2}+\cdots+D_{n-1})
\\
& =(z-\lambda )(z^{n-1}+(\lambda +D_1)z^{n-2}+\cdots+(\lambda ^{n-2}+D_1\lambda 
^{n-3}+\cdots+D_{n-2})\,z\,+ \\
& \,\,\,\,\,\,\,+\,(\lambda ^{n-1}+D_1\lambda ^{n-2}+\cdots+D_{n-1})).
\end{array}
$$  
\end{proposition}
\ \

The second lemma is a contribution 
of differential calculus to algebra.  

\begin{proposition}
\protect\label{calculus}
If $\,\lambda_1,\,\lambda_2,\,\ldots,\,\lambda_n$
are the complex roots of the nth-degree polynomial 
$\,p$ and $\,p'$ is its derivative polynomial, then 
$$
p'(z)=\sum_{k=1}^n \frac{p(z)}{z-\lambda_k}.
$$
{\bf Proof}: From $\,\,p(z)=d_0\,(z-\lambda_1)(z-\lambda_2)\cdots(z-\lambda_n)$,\, 
it follows 
$$
\begin{array}{l@{\hspace{.3em}}l}
p'(z)&=d_0\,(z-\lambda_2)(z-\lambda_3)\cdots(z-\lambda_n)
+d_0\,(z-\lambda_1)(z-\lambda_3)\cdots(z-\lambda_n)+\cdots \\
&\hspace*{1em}
+d_0\,(z-\lambda_1)(z-\lambda_2)\cdots(z-\lambda_{n-1}) 
=\fracc{p(z)}{z-\lambda_1}+\fracc{p(z)}{z-\lambda_2}+
\cdots+\fracc{p(z)}{z-\lambda_n}.
\end{array}
$$
\end{proposition}
\
\indent\par
Thus, we are able to get

\begin{theorem}[Newton's formulas \cite{Newton}]
\protect\label{TNewton}
If $\,\lambda_1,\,\lambda_2,\,\ldots,\,\lambda_n$
\,are the complex roots of the polynomial 
$\,p(z)=z^n+D_1\,z^{n-1}+D_2\,z^{n-2}+\cdots+D_{n-1}\,z+D_n$ 
\,with complex coefficients $\,D_k$ and 
$\,T_k=\lambda_1\mbox{}^k+\lambda_2\mbox{}^k+\cdots
+\lambda_n\mbox{}^k$ \,then
\begin{displaymath}
T_k+D_1\,T_{k-1}+\cdots+D_{k-1}\,T_1+k\,D_k=0,\quad 
k=1,\,2,\,\ldots,\,n.
\end{displaymath}
{\bf Proof}: Propositions \ref{seila} and \ref{calculus} yield 
$$
\begin{array}{l@{\hspace{.3em}}l@{\hspace{0em}}l}
\lefteqn{\,\,\,\,\,\,\,\, nz^{n-1}+(n-1)D_1z^{n-2}+(n-2)D_2z^{n-3}
+\cdots+D_{n-1}=} 
\\[1ex]
&=\sum\limits_{k=1}^n\bigg\{& z^{n-1}+(\lambda_k+D_1)z^{n-2}
+(\lambda_k^2+D_1\lambda_k+D_2)z^{n-3}+\cdots\\
&&+(\lambda_k^{n-1}+D_1\lambda_k^{n-2}+\cdots+D_{n-2}\lambda_k
+D_{n-1})\bigg\} \\
&\lefteqn{=nz^{n-1}+(T_1+nD_1)z^{n-2}+(T_2+D_1T_1+nD_2)z^{n-3}+
\cdots} \\
&\lefteqn{\hspace{1em}+(T_{n-1}+D_1T_{n-2}+\cdots
+D_{n-2}T_1+nD_{n-1}).}
\end{array}
$$
Equating coefficients we obtain 
$$
\begin{array}{l}
(n-1)D_1=T_1+nD_1, \\
(n-2)D_2=T_2+D_1T_1+nD_2, \\
\hspace*{5em}\vdots \\
D_{n-1}=T_{n-1}+D_1T_{n-2}+\cdots+D_{n-2}T_1+nD_{n-1},
\end{array}
$$
from which follow the $(n-1)$ first relations to be shown.  
Since each $\,\lambda_k$ is a root of $\,p$ it verifies 
$0=\lambda_k^n+D_1\lambda_k^{n-1}+\cdots+D_{n-1}\lambda_k+D_n$.  
The addition of these relations yields 
$$
0=\sum\limits_{k=1}^n\left\{\lambda_k^n+D_1\lambda_k^{n-1}+\cdots
+D_{n-1}\lambda_k+D_n\right\}=T_n+D_1T_{n-1}+\cdots+D_{n-1}T_1+nD_n,
$$
which is the remaining formula to be proved.
\end{theorem}

\section{The trace formulas}
\protect\label{TFormulas}

\indent\par
The characteristic polynomial \cite{charpoly} 
of a complex linear operator 
$\,\T$ in a finite dimensional complex vector space $\V$ is defined 
by $\,p(z)=\det\,(z\unity-\T)$, 
where $\,\unity$ is the identity operator in $\V$.  
The degree of $\,p$ as a polynomial in the complex variable $\,z$ 
equals the dimension of $\V$ as a complex vector space; 
its roots are called the characteristic values of $\,\T$ 
and the set of characteristic values is called the spectrum of $\,\T$.  

These definitions and properties lead to 
\begin{proposition}
\protect\label{trace}
The trace of a complex linear operator $\,\T$ equals the sum of 
its characteristic values counted with multiplicities.  \\
{\bf Proof}: 
$\det\left(z\unity-\T\right)
=(z-\lambda_1)(z-\lambda_2)\cdots(z-\lambda_n)$.  
By the matrix representation $T^i\mbox{}_j$ of $\,\T$ we cast 
the left-hand side as $z^n-\left(T^1\mbox{}_1+T^2\mbox{}_2+\cdots
+T^n\mbox{}_n\right)z^{n-1}+$ terms of order $\leq n-2$.  
In the right-hand side we have $z^n-\left(\lambda_1+\lambda_2+
\cdots+\lambda_n\right)z^{n-1}+$ terms of order $\leq n-2$.  
Equating coefficients we obtain 
$$
T^1\mbox{}_1+T^2\mbox{}_2+\cdots+T^n\mbox{}_n=\lambda_1+\lambda_2+
\cdots+\lambda_n.
$$
\end{proposition}

The roots of unity 
in the complex field yield the following factorization lemma 
in the associative algebra of complex linear operators.  
\begin{proposition}
\protect\label{Tpow}
If $\,\T$ and $\,\W$ are commuting complex linear operators 
and $\,\theta={\rm e}^{i2\pi/k}=\cos{\frac{2\pi}{k}}
+i\,\sin{\frac{2\pi}{k}}$\, with $k$ a positive integer, then 
$$
\T^k-\W^k=(\T-\W)(\T-\theta\,\W)(\T-\theta^2\W)\cdots
(\T-\theta^{k-1}\W).
$$
{\bf Proof}: As $\,\T$ and $\,\W$ commute we collect terms as
$$
\begin{array}{l@{\hspace{3em}}l}
\lefteqn{(\T-\W)(\T-\theta\,\W)(\T-\theta^2\W)\cdots
(\T-\theta^{k-1}\W)=} \\
& =\T^k+d_1\W\T^{k-1}+d_2\W^2\T^{k-2}+\cdots+d_{k-1}\W^{k-1}\T+d_k\W^k,
\end{array}
$$
where the complex numbers $\,\,d_1,\,d_2,\,\ldots,\,d_k$\, are the 
coefficients of the polynomial 
$(z-1)(z-\theta)(z-\theta^2)\cdots(z-\theta^{k-1})=\,z^k-1$, since 
$\,\left(\theta^j\right)\mbox{}^k=
\left(\left({\rm e}^{i2\pi/k}\right)\mbox{}^j\right)\mbox{}^k=1$\, 
for each $\,j=0,\,1,\,\ldots,\,k-1$.  Therefore $\,d_1=d_2=\ldots=d_{k-1}=0$\,
 and $\,d_k=-1$.  
\end{proposition}
 From the multiplicative and homogeneity properties 
of determinants we get 
 
\begin{proposition}
\protect\label{lambdak}

If $\,\lambda_1,\,\lambda_2,\,\ldots,\,\lambda_n$ are the 
characteristic values of the complex linear operator $\,\T$ then 
$\,\lambda_1\mbox{}^k,\,\lambda_2\mbox{}^k,\,\ldots,\,
\lambda_n\mbox{}^k$ are the characteristic values 
of the linear operator $\,\T^k$.  \\
 
{\bf Proof}: For the particular case where $\,\W=w\unity$, 
proposition \ref{Tpow} yields 
\begin{displaymath}
\begin{array}{r@{\hspace{.3em}}l}
&\,\,\,\,\,\,\,\,\,\det\left(\T^k-w^k\unity\right)=\det\left(\T-w\unity\right)
\det\left(\T-\theta w\unity\right)\cdots
\det\left(\T-\theta^{k-1}w\unity\right)\\
&\,\,=(\lambda_1-w)\cdots(\lambda_n-w)\,(\lambda_1-\theta w)\cdots
(\lambda_n-\theta w)\cdots(\lambda_1-\theta^{k-1}w)\cdots
(\lambda_n-\theta^{k-1}w) \\
&\,\,=(\lambda_1-w)(\lambda_1-\theta w)\cdots
(\lambda_1-\theta^{k-1}w)\cdots(\lambda_n-w)(\lambda_n-\theta w)
\cdots(\lambda_n-\theta^{k-1}w) \\
&\,\,=(\lambda_1\mbox{}^k-w^k)\cdots(\lambda_n\mbox{}^k-w^k).
\end{array}
\end{displaymath}
Thus $\,\,\det\left(z\unity-\T^k\right)=(z-\lambda_1\mbox{}^k)\cdots(z-\lambda_n\mbox{}^k).$
\end{proposition}
\  
\par Collecting the former results we are ready to achieve 
the {\bf recursive formulas} enclosed by 
\begin{theorem}
\protect\label{powtrace}
If $\,p(z)=z^n+D_1z^{n-1}+\cdots+D_{n-1}z+D_n$ 
is the characteristic polynomial of the complex linear operator 
$\,\T$ and $\,T_k=\mbox{\rm trace}\,(\T^k)$ \,then
\begin{displaymath}
T_k+D_1\,T_{k-1}+\cdots+D_{k-1}\,T_1+k\,D_k=0,\quad 
k=1,\,2,\,\ldots,\,n.
\end{displaymath}
{\bf Proof}: It is a straightforward consequence 
of theorem \ref{TNewton} 
together with propositions \ref{trace} and \ref{lambdak}.\
\end{theorem}

The key content of theorem \ref{powtrace} shall be stood out 
by the {\bf trace formulas statement:} {\bf the coefficients of the characteristic polynomial 
of a linear operator can be recursively computed 
as polynomial functions in the traces 
of its successive powers.}\par 
The {\bf trace formulas }
{\large $D_k=D_k(T_1,\,T_2,\,\ldots,\,T_{k-1},\,T_k)$}\,
for the coefficients of the characteristic polynomial are 
listed below for $k$ up to 8.\  
\begin{equation}
\protect\label{tformulas}
\begin{array}{l@{\hspace{.3em}}l}
D_1=&-T_1,  \\[1ex]
D_2=&\frac{1}{2}T_1\mbox{}^2-\frac{1}{2}T_2, \\[1ex]
D_3=&-\frac{1}{6}T_1\mbox{}^3+\frac{1}{2}T_1T_2-\frac{1}{3}T_3 ,\\[1ex]
D_4=&\frac{1}{24}T_1\mbox{}^4-\frac{1}{4}T_1\mbox{}^2T_2
+\frac{1}{3}T_1T_3+\frac{1}{8}T_2\mbox{}^2-\frac{1}{4}T_4 ,\\[1ex]
D_5=&-\frac{1}{120}T_1\mbox{}^{5}+\frac{1}{12}T_1\mbox{}^{3}T_2- 
\frac{1}{6}T_1\mbox{}^{2}T_3-\frac{1}{8}T_1T_2\mbox{}^{2}+ 
\frac{1}{4}T_1T_4+\frac{1}{6}T_2T_3-\frac{1}{5}T_5 ,\\[1ex]
D_6=&\frac{1}{720}T_1\mbox{}^{6}-\frac{1}{48}T_1\mbox{}^{4}T_2
+\frac{1}{18}T_1\mbox{}^{3}T_3+\frac{1}{16}T_1\mbox{}^2T_2\mbox{}^{2}
-\frac{1}{8}T_1\mbox{}^2T_4-\frac{1}{6}T_1T_2T_3
 \\[1ex]
&+\frac{1}{5}T_1T_5-\frac{1}{48}T_2\mbox{}^{3}+\frac{1}{8}T_2T_4+
\frac{1}{18}T_3\mbox{}^{2}-\frac{1}{6}T_6 , \\[1ex]
D_7=&-\frac{1}{5040}T_1\mbox{}^7+\frac{1}{240}T_1\mbox{}^{5}T_2
-\frac{1}{72}T_1\mbox{}^4T_3-\frac{1}{48}T_1\mbox{}^3T_2\mbox{}^2
+\frac{1}{24}T_1\mbox{}^3T_4+\frac{1}{12}T_1\mbox{}^2T_2T_3\\[1ex]
&-\frac{1}{10}T_1\mbox{}^2T_5
+\frac{1}{48}T_1T_2\mbox{}^3-\frac{1}{8}T_1T_2T_4
-\frac{1}{18}T_1T_3\mbox{}^2+\frac{1}{6}T_1T_6 -\frac{1}{24}T_2\mbox{}^2T_3\\[1ex]
&+\frac{1}{10}T_2T_5
+\frac{1}{12}T_3T_4-\frac{1}{7}T_7 ,\\[1ex]
D_8=&\frac{1}{40320}T_1\mbox{}^8-\frac{1}{1440}T_1\mbox{}^6T_2
+\frac{1}{360}T_1\mbox{}^5T_3
+\frac{1}{192}T_1\mbox{}^4T_2\mbox{}^2
-\frac{1}{96}T_1\mbox{}^4T_4\\[1ex]
& -\frac{1}{36}T_1\mbox{}^3T_2T_3+\frac{1}{30}T_1\mbox{}^3T_5
-\frac{1}{96}T_1\mbox{}^2T_2\mbox{}^3+\frac{1}{16}T_1\mbox{}^2T_2T_4
+\frac{1}{36}T_1\mbox{}^2T_3\mbox{}^2
\\[1ex]
&-\frac{1}{12}T_1\mbox{}^2T_6+
\frac{1}{24}T_1T_2\mbox{}^2T_3-\frac{1}{10}T_1T_2T_5-\frac{1}{12}T_1T_3T_4+
\frac{1}{7}T_1T_7+\frac{1}{384}T_2\mbox{}^4\\[1ex]
&-\frac{1}{32}T_2\mbox{}^2T_4
-\frac{1}{36}T_2T_3\mbox{}^2+\frac{1}{12}T_2T_6+\frac{1}{15}T_3T_5+
\frac{1}{32}T_4\mbox{}^2-\frac{1}{8}T_8. \\[3ex] 
\end{array}
\end{equation}
 The {\bf characteristic formulas} for dimensions up to four are written down:
\begin{eqnarray}
\protect\label{determinants}
\det \, \left( z\unity-\T \right) & = & z-T_1, \nonumber \\
\det\,\left(z\unity-\T\right) & = &
z^2-T_1z+\frac{1}{2}(T_1\mbox{}^2-T_2),\nonumber  \\
\det\,\left(z\unity-\T\right) & = &
z^3-T_1z^2+\frac{1}{2}(T_1\mbox{}^2-T_2)z
-{\frac{1}{6}}(T_1\mbox{}^3-3T_1T_2+2T_3),  \\
\det\,\left(z\unity-\T\right) & = &
z^4-T_1z^3+\frac{1}{2}(T_1\mbox{}^2-T_2)z^2
-{\frac{1}{6}}(T_1\mbox{}^3-3T_1T_2+2T_3)z+ \nonumber \\
&  & \,\,\,\,\,\,\,\,+ {\frac{1}{24}} (T_1\mbox{}^4-6T_1\mbox{}^2T_2+8T_1T_3
+3T_2\mbox{}^2-6T_4). \nonumber 
\end{eqnarray}

\section{Null identities}
\protect\label{Cayley}

\indent\par
The polynomial $\,p(z)=d_0z^n+d_1z^{n-1}+\cdots+d_{n-1}z+d_n$ 
with complex coefficients $\,d_k$ is said to annihilate 
the complex linear operator $\,\T$ if 
$\,p(\T)=d_0\T^n+d_1\T^{n-1}+\cdots+d_{n-1}\T+d_n\unity=\0$, 
the identically null operator.  

The knowledge of the principal ideal \cite{annulet} 
of polynomials that annihilate 
a linear operator $\,\T$ is essential 
to attain computational skill 
in the associative algebra generated by $\,\T$.  
To ensure this goal we state without proof 
one of the fundamental results in linear algebra.  

\begin{theorem}[Cayley-Hamilton \cite{Cayley}]
\protect\label{TCayley}
The characteristic polynomial 
of a linear operator 
annihilates it.  
\end{theorem}

\par Joining Cayley-Hamilton's theorem 
with trace formulas statement 
we conclude that {\bf for each finite dimension 
of the underlying vector space 
there is a fundamental null identity 
in the associative algebra of linear operators.} For instance, 
in dimensions up to four, the 
characteristic formulas (\ref{determinants}) 
yield the following {\bf null identities:} 
\ \\ $$
\begin{array}{l@{\hspace{0pt}}c@{\hspace{0pt}}c%
               @{\hspace{0pt}}c@{\hspace{-.2em}}r}
\T&-\hfill T_1\unity\hfill&\makebox[0pt][l]{$=\0 ,$}
\hfill \\[2ex]
\T^2&-\hfill T_1\T\hfill
&+\hfill\fracc{1}{2}(T_1\mbox{}^2-T_2)\unity\hfill&
\makebox[0pt][l]{$=\0 ,$}\hfill \\[2ex]
\T^3&-T_1\T^2&+\hfill\fracc{1}{2}(T_1\mbox{}^2-T_2)\T
\hfill&-\fracc{1}{6}(T_1\mbox{}^3-3T_1T_2+2T_3)\unity=\0 ,
\hfill \\[2ex]
\T^4&-T_1\T^3&+\fracc{1}{2}(T_1\mbox{}^2-T_2)\T^2
&-\fracc{1}{6}(T_1\mbox{}^3-3T_1T_2+2T_3)\T+ \hfill\\[2ex]
&&&&\makebox[0pt][r]
{$+\fracc{1}{24}(T_1\mbox{}^4-6T_1\mbox{}^2T_2
+8T_1T_3+3T_2\mbox{}^2-6T_4)\unity=\0 .$}
\end{array}
$$
 
\section{The finite-dimensional resolvent kernel}
\protect\label{Kernel}

\indent\par The resolvent of a complex linear operator $\,\T$ 
is the operator-valued function $\,\R$ defined by 
$\,\R(z)=\left(z\unity-\T\right)^{-1}$.  
It is a well-known fact that $\,\R$ is an operator-valued analytic    
function outside the spectrum of $\,\T$.  
We shall refine such a result showing that in the finite dimensional case  $\,\R$ is an operator-valued rational function 
completely tied by a finite set of complex-valued rational functions, 
whose coefficients are exactly the trace formulas.  

In this vein we need an improved version 
of the remainder theorem in the associative algebra 
of complex linear operators, whose content is 
the extension of proposition \ref{seila} 
to operator-valued polynomials.  

\begin{proposition}
\protect\label{remainder}
If $\,p(w)=w^n+D_1w^{n-1}+\cdots+D_{n-1}w+D_n$ 
is a polynomial with complex coefficients $\,D_k$ 
and $\,\T$ and $\,\W$ are commuting complex linear operators 
then
$$
\begin{array}{l@{\hspace{.3em}}l}
p(\W)-p(\T)=(\W-\T)&\left[\T^{n-1}+(\W+D_1\unity)\T^{n-2}
+(\W^2+D_1\W+D_2\unity)\T^{n-3}+ \right.\\[1ex]
&\left.\hspace{.3em}
+\cdots+(\W^{n-1}+D_1\W^{n-2}+\cdots+D_{n-1}\unity)
\unity\right].
\end{array}
$$
{\bf Proof}: By induction on the degree.  For $n=1$,\,\,\,
$p(\W)-p(\T)=(\W+D_1\unity)-(\T+D_1\unity)=\W-\T$.  
For generic $n$,\,\,\, $p(\W)-p(\T)=$
$$
\begin{array}{l@{\hspace{.3em}}l@{\hspace{.2em}}l}
&\lefteqn{
=\W(\W^{n-1}+D_1\W^{n-2}+\cdots+D_{n-1}\unity)
+D_n\unity +}\\[1ex]
&\lefteqn{
\hspace{1.4em}-\T(\T^{n-1}+D_1\T^{n-2}+\cdots
+D_{n-1}\unity)-D_n\unity }\\[1ex]
&\lefteqn{=(\W-\T)(\W^{n-1}+D_1\W^{n-2}+\cdots
+D_{n-1}\unity) +}\\[1ex]
&+\T\bigg[&(\W^{n-1}+D_1\W^{n-2}+\cdots+D_{n-1}\unity) +\\[1ex]
&&-(\T^{n-1}+D_1\T^{n-2}+\cdots+D_{n-1}\unity)\bigg].
\end{array}
$$
From the inductive hypothesis the last expression reads
$$
\begin{array}{l@{\hspace{.2em}}l@{\hspace{.2em}}l}
(\W-\T)&\lefteqn{
(\W^{n-1}+D_1\W^{n-2}+\cdots+D_{n-1}\unity) }\\[1ex]
&+\T(\W-\T)&\left[\T^{n-2}+(\W+D_1\unity)\T^{n-3}+\cdots 
\right.\\[1ex]
&&\hspace{.5em}\left.+(\W^{n-2}+D_1\W^{n-3}+\cdots
+D_{n-2}\unity)\unity\right] \\[1ex]
&\lefteqn{
=(\W-\T)\left[\T^{n-1}+(\W+D_1\unity)\T^{n-2}+\cdots\right. 
+(\W^{n-2}+D_1\W^{n-3}+ }\\[1ex]
&&\left.+\cdots+D_{n-2}\unity)\T
+(\W^{n-1}+D_1\W^{n-2}+\cdots+D_{n-1}\unity)\unity\right].
\end{array}
$$
\end{proposition}
\
\par We are ready to achieve a {\bf rational formula} 
for the resolvent kernel.  

\begin{theorem}
\protect\label{resolvent}
If $\,p(w)=w^n+D_1w^{n-1}+\cdots+D_{n-1}w+D_n$ 
is the characteristic polynomial 
of the complex linear operator $\,\T$ 
and $\,z$ is any complex number 
not belonging to the spectrum of $\,\T$ then 
$$
\begin{array}{l@{\hspace{.3em}}l}
\left(z\unity-\T\right)^{-1}
=&\fracc{1}{p(z)}\T^{n-1}
+\fracc{z+D_1}{p(z)}\T^{n-2}
+\fracc{z^2+D_1z+D_2}{p(z)}\T^{n-3}
+\cdots \\[2ex]
&+\fracc{z^{n-1}+D_1z^{n-2}+\cdots
       +D_{n-1}}{p(z)}\unity.
\end{array}
$$
{\bf Proof}: It is enough to consider 
Cayley-Hamilton's theorem for the linear operator $\,\T$ 
and set $\,\W=z\unity$ in proposition \ref{remainder}; notice that in such case 
$\,\W^k+D_1\W^{k-1}+\cdots+D_k\unity
=(z^k+D_1z^{k-1}+\cdots+D_k)\unity$ 
\,for each $k=1,\,2,\,\ldots,\,n$.  
\end{theorem}
\par The joint content of theorem \ref{resolvent} 
and the trace formulas statement 
shall be pointed out as {\bf: for each finite dimension 
of the underlying vector space 
there is a fundamental rational formula 
for the resolvent kernel of a linear operator.} 
For instance, in dimensions up to four, 
the trace formulas (\ref{tformulas}) provide the following {\bf resolvent formulas:}

\begin{equation}
\protect\label{upto4}
\hspace*{-3em}
\begin{array}{l@{\hspace{.3em}}l}
\lefteqn{\left(z\unity-\T\right)^{-1}
=(z-T_1)^{-1}\unity ,}\\[1ex]
\lefteqn{\left(z\unity-\T\right)^{-1}
=\left(z^2-T_1z+\frac{1}{2}(T_1\mbox{}^2-T_2)\right)^{-1}
\left[\nul\T+(z-T_1)\unity\right] ,}\\[1.5ex]
\left(z\unity-\T\right)^{-1}=\bigg(z^3 \nul 
& \lefteqn{ -T_1z^2+\frac{1}{2}(T_1\mbox{}^2-T_2)z
-\fracc{1}{6}(T_1\mbox{}^3-3T_1T_2+2T_3)\bigg)^{-1}
\bigg[\T^2 +}\\[1.5ex]
&\lefteqn{ +(z-T_1)\T+\left(z^2-T_1z+\frac{1}{2}
(T_1\mbox{}^2-T_2)\right)\unity\bigg],}\hspace*{22em}\\[2ex]
\lefteqn{\left(z\unity-\T\right)^{-1}=
\left(z^4-T_1z^3+\frac{1}{2}(T_1\mbox{}^2-T_2)z^2
-\frac{1}{6}(T_1\mbox{}^3-3T_1T_2+2T_3)z +\right.}\\[2ex]
&\lefteqn{\left.+\frac{1}{24}(T_1\mbox{}^4-6T_1\mbox{}^2T_2
+8T_1T_3+3T_2\mbox{}^2-6T_4)\right)^{-1}
\bigg[\T^3 +}\\[2ex]
&\lefteqn{ +(z-T_1)\T^2+\left(z^2-T_1z
+\frac{1}{2}(T_1\mbox{}^2-T_2)\right)\T
+\left(z^3 \right.+}\\[2ex]
&\lefteqn{\left. -T_1z^2+\frac{1}{2}(T_1\mbox{}^2-T_2)z
-\frac{1}{6}(T_1\mbox{}^3-3T_1T_2+2T_3)
\right)\unity\bigg].}
\end{array}
\end{equation}\
\section{Applications to general relativity}
\protect\label{aplication}

\indent\par
In general relativity 
the fundamental physical object is an effective geometry, 
mathematically represented by a second-rank covariant tensor field $\mbox{\bf g}$, defined throughout a suitable 4-dimensional manifold.  
In a local coordinate system 
$\mbox{\bf x}=\left(x^\alpha\right)$ 
the metric tensor can be written as 
$\mbox{\bf g}=g_{\mu\nu}{\rm d}x^\mu\otimes{\rm d}x^\nu$.  

Investigations on general relativity 
are frequently carried out under the assumption 
that there exists some background geometry 
$\stackrel{o}{\mbox{\bf g}}=
{\stackrel{o}{g}\mbox{$\!$}_{\mu\nu}}
{\rm d}x^\mu\otimes{\rm d}x^\nu$.  
The metric properties to be assumed 
on $\stackrel{o}{\mbox{\bf g}}$ 
vary depending on the gravitational scenario.  
In order to perform calculations it suffices 
to set $\stackrel{o}{\mbox{\bf g}}$ 
a Ricci-flat metric.  
However, to interpret the results as physically meaningful 
it is generally agreed that one should require 
$\stackrel{o}{\mbox{\bf g}}$ to be a flat metric.  
Even this case is sometimes thought to be too broad, 
as some authors claim to set harmonic coordinates \cite{Fock} 
or even cartesian coordinates \cite{Landau}.  
We do not take into account here any suplementary conditions 
on the background geometry $\stackrel{o}{\mbox{\bf g}}$.

The effective and background geometries are related 
by a tensor field 
$\mbox{\bf h}=h_{\mu\nu}{\rm d}x^\mu\otimes{\rm d}x^\nu$ 
by a connecting equation, 
the generally accepted form of which being 
$\mbox{\bf g}={\stackrel{o}{\mbox{\bf g}}}+\mbox{\bf h}$.  
The scalar density $\sqrt{-g}d{\rm x}^4$ 
associated with $\mbox{\bf g}$ requires us to compute 
$\det\,(\mbox{\bf g})
=\det\,(\stackrel{o}{\mbox{\bf g}}+\mbox{\bf h})$.  
This can be achieved by means of 
\begin{proposition}
If $\mbox{\bf g}
=\stackrel{o}{\mbox{\bf g}}\mbox{$\!$}+\mbox{\bf h},\,\VH
=\stackrel{o}{\mbox{\bf g}}\mbox{$\!$}^{-1}\mbox{\bf h},\,
H_k={\rm trace}\,\left(\VH^k\right)$ 
and the underlying manifold is 4-dimensional then 
$$
\frac{\det\,(\mbox{\bf g})}{\det\,(\stackrel{o}{\mbox{\bf g}})}
=\begin{array}[t]{l@{\hspace{.2em}}l}
1+H_1&+\fracc{1}{2}(H_1\mbox{}^2-H_2)
+\fracc{1}{6}(H_1\mbox{}^3-3H_1H_2+2H_3) +\\[2ex]
&+\fracc{1}{24}(H_1\mbox{}^4-6H_1\mbox{}^2H_2
+8H_1H_3+3H_2\mbox{}^2-6H_4) .
\end{array}
$$
{\bf Proof}: From 
$\mbox{\bf g}=\stackrel{o}{\mbox{\bf g}}\mbox{$\!$}+\mbox{\bf h}
=\stackrel{o}{\mbox{\bf g}}\left(\unity+\VH\right)$ 
it follows that 
$\det\,(\mbox{\bf g})
=\det\,(\stackrel{o}{\mbox{\bf g}})\det\,(\unity+\VH)$.  
Now it is enough to consider 
the 4-dimensional characteristic formula in (\ref{determinants}) 
with $z=1$ and $\T=-\VH$;  
notice that $\T^k=(-)^k\VH^k$ implies $T_k=(-)^kH_k$.  
\end{proposition}

The Levi-Civita connection associated with $\mbox{\bf g}$ 
requires us to compute 
$\mbox{\bf g}^{-1}
=\left({\stackrel{o}{\mbox{\bf g}}+\mbox{\bf h}}\right)^{-1}
=\left(\unity+\VH\right)^{-1}{\stackrel{o}{\mbox{\bf g}}
\mbox{$\!$}^{-1}}$.  
The known explicit form 
is given by the Neumann series \cite{Neumann} 
\begin{displaymath}
\left(\unity+\VH\right)^{-1}=\unity-\VH+\VH^2-\VH^3+\cdots
\end{displaymath}
In local coordinates it reads 
\begin{displaymath}
g^{\mu\nu}=\left(\delta^\mu\mbox{}_\beta
-\stackrel{o}{g}\mbox{$\!$}^{\mu\alpha}h_{\alpha\beta}
+\stackrel{o}{g}\mbox{$\!$}^{\mu\alpha}h_{\alpha\lambda}
\stackrel{o}{g}\mbox{$\!$}^{\lambda\epsilon}h_{\epsilon\beta}
-\stackrel{o}{g}\mbox{$\!$}^{\mu\alpha}h_{\alpha\lambda}
\stackrel{o}{g}\mbox{$\!$}^{\lambda\epsilon}h_{\epsilon\rho}
\stackrel{o}{g}\mbox{$\!$}^{\rho\sigma}h_{\sigma\beta}+\cdots
\right)\stackrel{o}{g}\mbox{$\!$}^{\beta\nu},
\end{displaymath}
where $g^{\mu\nu}$ and $\stackrel{o}{g}\mbox{$\!$}^{\mu\nu}$ 
are well-defined by 
$g^{\mu\alpha}g_{\alpha\nu}=\delta^\mu\mbox{}_\nu
=\stackrel{o}{g}\mbox{$\!$}^{\mu\beta}
 \stackrel{o}{g}\mbox{$\!$}_{\beta\nu}$.  

Besides convergence requirements on the above series, 
we stress that such expression leads to technical difficulties 
when developing the Lagrangian variational formalism.  
These problems were already dealt with in the literature, 
the proposed solution being to modify 
the above form of the connecting equation \cite{Grishchuk}.  
We show how to overcome 
such drawbacks by means of 

\begin{proposition}
\protect\label{GR}
If $\mbox{\bf g}
=\stackrel{o}{\mbox{\bf g}}\mbox{$\!$}+\mbox{\bf h},\,\VH
=\stackrel{o}{\mbox{\bf g}}\mbox{$\!$}^{-1}\mbox{\bf h},\,
H_k={\rm trace}\,\left(\VH^k\right)$ 
and the underlying manifold is 4-dimensional then 
\begin{displaymath}
\hspace*{-1em}
\begin{array}{l@{\hspace{.2em}}l}
\mbox{\bf g}^{-1}=\left(\nul\right.&
1+H_1+\fracc{1}{2}(H_1\mbox{}^2-H_2)
+\fracc{1}{6}(H_1\mbox{}^3-3H_1H_2+2H_3) +\\[2ex]
&\left.+\fracc{1}{24}(H_1\mbox{}^4-6H_1\mbox{}^2H_2
+8H_1H_3+3H_2\mbox{}^2-6H_4)\right)^{-1}\bigg[-\VH^3 +\\[2ex]
&\hspace{4em}
+(1+H_1)\VH^2-\left(1+H_1+\fracc{1}{2}(H_1\mbox{}^2-H_2)\right)\VH
+\left(1 \nul\right.+\\[2ex]
&\hspace{4em}
+H_1+\fracc{1}{2}(H_1\mbox{}^2-H_2)\left.
+\fracc{1}{6}(H_1\mbox{}^3-3H_1H_2+2H_3)\right)\unity\bigg]
\stackrel{o}{\mbox{\bf g}}\mbox{$\!$}^{-1}.
\end{array}
\end{displaymath}
{\bf Proof}: From 
$\mbox{\bf g}=\stackrel{o}{\mbox{\bf g}}\mbox{$\!$}+\mbox{\bf h}
=\stackrel{o}{\mbox{\bf g}}\left(\unity+\VH\right)$ 
it follows that ${\mbox{\bf g}^{-1}}
=\left(\unity+\VH\right)^{-1}
{\stackrel{o}{\mbox{\bf g}}\mbox{$\!$}^{-1}}$.  
Now it is enough to consider 
the 4-dimensional resolvent formula in (\ref{upto4}) 
with $z=1$ and $\T=-\VH$;  
notice that $\T^k=(-)^k\VH^k$ implies $T_k=(-)^kH_k$.  
\end{proposition}

\section*{Acknowledgements}

\indent\par
The author thanks Dr. Renato Klippert Barcellos for having collaborated with him in this work, for valuable discussions, and for the implementation of the computer algebra algorithm. The author owes special thanks to Dr. Nelson Pinto Neto and Dr. Vitorio Alberto De Lorenci for encouragement, and for having drawn the attention of him to the relevance of the subject of the present work.  The author is also indebted to Carla da Fonseca-Bernetti for helpful corrections of the manuscript.  

This work was financially supported by CNPq.


\end{document}